# Normal stress effects in the gravity driven flow of granular materials

Wei-Tao Wu[1], Nadine Aubry[2], James F. Antaki[1], Mehrdad Massoudi[3*]

1. Department of Biomedical Engineering, Carnegie Mellon University,
Pittsburgh, PA, 15213, USA
2. Department of Mechanical and Industrial Engineering, Northeastern University,
Boston, MA 02115, USA
3. U. S. Department of Energy, National Energy Technology Laboratory (NETL),
Pittsburgh, PA, 15236, USA

Address for correspondence:

Mehrdad Massoudi, PhD,

U. S. Department of Energy,

National Energy Technology Laboratory (NETL),

626 Cochrans Mill Road, P.O. Box 10940,

Pittsburgh, PA. 15236.

Email: Mehrdad.Massoudi@NETL.DOE.GOV

**Abstract**

In this paper, we study the fully developed gravity-driven flow of granular materials between two inclined planes. We assume that the granular materials can be represented by a modified form of the second-grade fluid where the viscosity depends on the shear rate and volume fraction and the normal stress coefficients depend on the volume fraction. We also propose a new isotropic (spherical) part of the stress tensor which can be related to the compactness of the (rigid) particles. This new term ensures that the rigid solid particles cannot be compacted beyond a point, namely when the volume fraction has reached the critical/maximum packing value. The numerical results indicate that the newly proposed stress tensor has an obvious and physically meaningful effects on both the velocity and the volume fraction fields.

## 1. Introduction

Granular materials are discrete solid macroscopic particles of different shapes and sizes with interstices filled with a fluid. Granular materials occur in many natural processes, such as the flow of rock, sand, snow and ice. In industrial applications, water and granular materials (grains, powders, coals, etc.[1]) are the first and the second mostly used materials. A granular medium does not behave as a classical solid continuum since it deforms to (takes) the shape of the vessel containing it; it is not exactly like a liquid, even though it can flow, for it can be piled into heaps; and it is not a gas since it will not expand to fill the container. In many ways the bulk solids resemble non-Newtonian (non-linear) fluids [2]. The behavior of granular materials, in general, is determined by interparticle cohesion, friction, collisions, etc [3]. A granular medium includes granular powders and granular solids with components ranging in size from about 10 μm up to 3 mm. According to many researchers, a powder is composed of particles up to 100 μm (diameter) with further subdivision into ultrafine (0.1 to 1.0 μm), superfine (1 to 10 μm), or granular (10 to 100 μm) particles, whereas a granular solid consists of particles ranging from about 100 to 3,000 μm [Brown and Richards (1970) [4]].

Granular materials present a multi-disciplinary field; they can be studied from different perspectives. For example, in order to characterize their rheological behavior, one can study the mechanics (or physics) of these complex materials by performing experiments, which are oftentimes very complicated. Recent review articles by Savage (1984) [5], Hutter and Rajagopal (1994) [6], and de Gennes (1998) [7], and books by Mehta (1994) [8], Duran (2000) [9], and Antony et al. (2004) [10] point to many of the important issues in modeling granular materials. From a theoretical perspective, there are two distinct, yet related methods that can be used: the statistical theories and the continuum theories. There are many recent review articles discussing the statistical theories [Herrmann (1999) [11], Herrmann and Luding (1998) [12]], and the kinetic theories as applied to granular materials [Goldhirsch (2003) [13] and Boyle and Massoudi (1990) [14]].

In this paper, we model the granular materials as a single phase continuum, ignoring the effects of the interstitial fluid. However, it should be pointed out that in many applications, the effect of the interstitial fluid is important and the problem should be studied using a multi-component method [see Rajagopal and Tao (1995) [15], Massoudi (2010) [16]]. In section 2, we will present the governing equations. In section 3, we will discuss the constitutive equations. In section 4, we introduce the geometry and kinematics of the problem. In section 5, we discuss the numerical results through a parametric study of the non-linear ordinary equations for the gravity driven flow of the granular materials between two inclined plates.

## 2. Governing Equations

We model the granular materials as a single component non-linear fluid. That is, we ignore the presence of the interstitial fluid and assume that the assembly of the densely-packed particles form a continuum. Let $\boldsymbol{X}$ denote the position of this continuum body. The motion can be represented as

$$\boldsymbol{x} = \chi(\boldsymbol{X}, t) \tag{1}$$

while the kinematical quantities associated with this motion are

$$\boldsymbol{v} = \frac{d\boldsymbol{x}}{dt} \tag{2}$$

$$\boldsymbol{D} = \frac{1}{2}\left(\frac{\partial \boldsymbol{v}}{\partial \boldsymbol{x}} + \left(\frac{\partial \boldsymbol{v}}{\partial \boldsymbol{x}}\right)^T\right) \tag{3}$$

where $\boldsymbol{v}$ is the velocity field, $\boldsymbol{D}$ is the symmetric part of velocity gradient, and $\frac{d}{dt}$ denotes differentiation with respect to time holding $\boldsymbol{X}$ fixed and superscript '$T$' designates the transpose of a tensor. The bulk density field, $\rho$, is

$$\rho = (1-\phi)\rho_0 \tag{4}$$

where $\rho_0$ is the pure density of granular materials, in the reference configuration; $\phi$ is the volume fraction, where $0 \leq \phi < \phi_{max} < 1$. The function $\phi$ is represented as a continuous function of position and time; in reality, $\phi$ is either one or zero at any position and at any given time. That is, in a sense, we have performed a homogenization process (see Collins(2005) [17]) whereby the shape and sizes of the particles in this idealized body have disappeared except through the presence of the volume fraction. For details see Massoudi (2001) [18] and Massoudi and Mehrabadi (2001) [19]. In reality, $\phi$ is never equal to one; its maximum value, generally designated as the maximum packing fraction, depends on the shape, size, method of packing, etc.

Having defined the basic kinematical parameters, we now look at the conservation equations. In the absence of any thermo-chemical and electro-magnetic effects, the governing equations for the flow of a single-component material are the conservation equations for mass, linear momentum, and angular momentum [20]. As we are only considering a purely mechanical system, the energy equation and the entropy inequality are not considered.

*Conservation of mass*

The conservation of mass is:

$$\frac{\partial \rho}{\partial t} + div(\rho \boldsymbol{v}) = 0 \tag{5}$$

where $\partial/\partial t$ is the derivative with respect to time, $div$ is the divergence operator.

*Conservation of linear momentum*

Let $\boldsymbol{T}$ represent the Cauchy stress tensor for the granular materials. Then the balance of the linear momentum is:

$$\rho \frac{d\boldsymbol{v}}{dt} = div\boldsymbol{T} + \rho \boldsymbol{b} \tag{6}$$

where $\frac{d\boldsymbol{v}}{dt} = \frac{\partial \boldsymbol{v}}{\partial t} + (grad\boldsymbol{v})\boldsymbol{v}$ and $\boldsymbol{b}$ stands for the body force.

*Conservation of Angular Momentum*

$$\boldsymbol{T} = \boldsymbol{T}^T \tag{7}$$

The above equation implies that in absence of couple stresses the Cauchy stress tensor is symmetric. The constitutive relation for the stress tensor need to be specified before any problem can be solved. In the next section, we will discuss this issue.

## 3. Constitutive Equation: The stress tensor

Most granular materials exhibit two unusual and peculiar characteristics: (i) normal stress differences, and (ii) yield criterion[1]. Reynolds (1885) [21] observed that in a bed of closely packed particles, the bed must expand if a shearing motion is to occur; this happens in order to increase the volume of the voids. Reynolds (1886) [22] called this phenomenon 'dilatancy' and he was able to describe the capillary action in wet sand. The concept of dilatancy, which in a larger context, is related to the normal stress differences in non-linear materials, is related the expansion of the void volumes in a packed granular arrangement when it is subjected to a deformation. This can be explained for an idealized case: in a bed of closely packed spheres, for a shearing motion to occur, for example, between two flat plates, the bed must expand by increasing its void volume. Reiner (1945, 1948) [23,24] used a non-Newtonian model to predict dilatancy in wet sand [see Massoudi (2011, 2013) [25,26]]. Perhaps the simplest constitutive equation which can describe the normal stress effects in non-linear fluids (related to phenomena such as 'die-swell' and 'rod-climbing', which are manifestations of the stresses that develop orthogonal to planes of shear) is the second-grade fluid, or the Rivlin-Ericksen fluid of grade two [Rivlin and Ericksen (1955) [27], Truesdell and Noll (1992) [28]]. This model is a special case of fluids of differential type [Dunn and Rajagopal (1995) [29]]. For a second-grade fluid, the Cauchy stress tensor is given by [30]:

$$\boldsymbol{T} = -p\boldsymbol{I} + \mu \boldsymbol{A}_1 + \alpha_1 \boldsymbol{A}_2 + \alpha_2 \boldsymbol{A}_1^2 \qquad (8)$$

where $p$ is the indeterminate part of the stress due to the constraint of incompressibility. $\mu$ is the coefficient of viscosity which may depends on shear rate, volume fraction, pressure, temperature, etc. [31–33], and $\alpha_1$ and $\alpha_2$ are material moduli which are commonly referred to as the normal stress coefficients [34]. The kinematical tensors $\boldsymbol{A}_1$ and $\boldsymbol{A}_2$ are defined through

$$\begin{cases} \boldsymbol{A}_1 &= \boldsymbol{L} + \boldsymbol{L}^T \\ \boldsymbol{A}_2 &= \dfrac{d\boldsymbol{A}_1}{dt} + \boldsymbol{A}_1 \boldsymbol{L} + \boldsymbol{L}^T \boldsymbol{A}_1 \\ \boldsymbol{L} &= grad\ \boldsymbol{v} \end{cases} \qquad (9)$$

---

[1] In this paper, we do not discuss the yield stress.

where **L** is the velocity gradient. The thermodynamics and stability of fluids of second grade have been studied in detail by Dunn and Fosdick (1974) [35]. They concluded that if the fluid is to be thermodynamically consistent in the sense that all motions of the fluid meet the Clausius-Duhem inequality and that the specific Helmholtz free energy be a minimum in equilibrium, then

$$\mu \geq 0 \qquad (10)\text{a}$$

$$\alpha_1 \geq 0 \qquad (10)\text{b}$$

$$\alpha_1 + \alpha_2 = 0 \qquad (10)\text{c}$$

It is known that for many non-linear fluids which are assumed to follow Equation (8), the experimental values reported for $\alpha_1$ and $\alpha_2$ do not satisfy the restriction[2] of equations (10)b and (10)c. For further details on this and other relevant issues in fluids of differential type, we refer the reader to the review article by Dunn and Rajagopal (1995) [29].For some applications, such as flow of ice or coal slurries, where the fluid is known to be shear-thinning (or shear-thickening), modified (or generalized) forms of the second-grade fluid have been proposed [see Man (1992) [36], Massoudi and Vaidya (2008) [37], Man and Massoudi (2010) [38]].

In this paper, we assume that the (flowing) granular materials can be modeled as a non-linear fluid (of second-grade type) capable of exhibiting normal stress effects, where the shear viscosity depends on the volume fraction and the shear rate and the normal stress coefficients depend on the volume fraction:

$$\boldsymbol{T} = \beta_0 \boldsymbol{I} + \mu(\phi, \dot{\gamma})\boldsymbol{A}_1 + \alpha_1(\phi)\boldsymbol{A}_2 + \alpha_2(\phi)\boldsymbol{A}_1^2 \qquad (11)$$

Here, $\beta_0$ is the spherical (isotropic part of the) stress tensor (which includes the pressure), $\mu$ is shear the viscosity and $\dot{\gamma} = \sqrt{1/2\text{tr}(\boldsymbol{A}_1^2)}$ is the shear rate, $\alpha_1$ and $\alpha_2$ are material moduli, which are commonly referred to as the normal stress coefficients. For these coefficients we assume:

$$\begin{cases} \mu &= \mu_r(\phi + \phi^2) \\ \alpha_1 &= \alpha_{10}(\phi + \phi^2) \\ \alpha_2 &= \alpha_{20}(\phi + \phi^2) \end{cases} \qquad (12)$$

---

[2] In an important paper, Fosdick and Rajagopal (1979) [47] show that irrespective of whether $\alpha_1 + \alpha_2$ is positive, the fluid is unsuitable if $\alpha_1$ is negative.

These expressions can be viewed as the Taylor series approximation for the material parameters [Rajagopal, et al (1994)[39]]. The above equations ensure that the stress vanishes as $\phi \to 0$.

In general, $\beta_0$ is treated as the spherical (the isotropic) part of the stress tensor which includes terms related to the pressure. In this paper, we propose that once the volume fraction has reached a critical value, further compacting, which can be measured as the increase in the local particles concentration, would lead to an additional term related to the isotropic spherical stress preventing further increase in the local volume fraction[3]. Therefore, we suggest that $\beta_0$ can be decomposed into two parts: one term is to account for the mechanism related to the pressure, while the second term is to account for the compacting of the particles,

$$\beta_0 = (\beta_p + \beta_r(\phi))\phi \tag{13}$$

Based on our previous studies (Rajagopal and Massoudi (1990) [40]), we assume

$$\beta_p = \beta_{00} < 0 \tag{14}$$

as compression should lead to densification of the materials, see [40,41].

We further propose that $\beta_r(\phi)$ term can be given by:

$$\beta_r = h(\phi)C_r g(\phi) \tag{15}$$

$$h(\phi) = \begin{cases} 0, \phi < \phi_c \\ 1, \phi \geq \phi_c \end{cases} \tag{16}$$

where $\phi_c$ is the critical value of the volume fraction; where $\phi \geq \phi_c$ the term $\beta_r$ appears in the equation. The values of $\phi_c$ may depend on various factors, such as the shape and the size distributions of the particles. In this paper, we assume $g(\phi) = \phi$. $C_r$ is a material parameter related to how much the particles can be compacted. For very rigid particles, $C_r$ is large, ensuring that the volume fraction of the granular materials cannot be larger than $\phi_c$. Furthermore, for ease of numerical calculations, we replace $h(\phi)$ in equation (16) with a smooth step function, such that

$$h(\phi) = \frac{1}{1 + \exp(-2S(\phi - \phi_c))} \tag{17}$$

---

[3] When the particles are large and heavy, there is a tendency for them to settle under the action of gravity [48,49]. During the sedimentation process, the volume fraction of the particles reaches a critical value beyond which the particles cannot be compacted any further (due to the rigidity of the particles).

where $S$ is a parameter related to the slope of the step function. For example, if $S$ is chosen as 3000, then $h(\phi_c - 0.001)/h(\phi_c) < 0.01$ and $h(\phi_c - 0.005) \sim o(10^{-13})$, therefore $\beta_r$ is negligible when $\phi$ is close to $\phi_c$.

The viscosity is assumed to be modeled as a Carreau-type fluid viscosity, where the viscosity depends on the shear rate. When the shear rate is close to zero, the viscosity approaches a lower limit, $\mu_0$; when the shear rate is close to infinity, the viscosity approaches an upper limit, $\mu_\infty$. Following Yeleswarapu (1994)[42], we assume $\mu_r$ is given by:

$$\mu_r = \mu_\infty + (\mu_0 - \mu_\infty)\frac{1 + \ln(1 + k\dot{\gamma})}{1 + k\dot{\gamma}} \tag{18}$$

where $\mu_0$ and $\mu_\infty$ are the viscosities when the shear rate approaches zero and infinity, respectively, and $k$ is the shear thinning parameter. Different types of solid particles can have different functions/expressions for $\mu_0$, $\mu_\infty$ and $k$.

## 4. Geometry and the kinematics of the flow

The geometry and the kinematics of the flow are shown in Figure 1. We consider a gravity driven flow between two inclined flat plates. This flow configuration has been studied extensively in the context of non-linear fluids and granular [25,43–45].

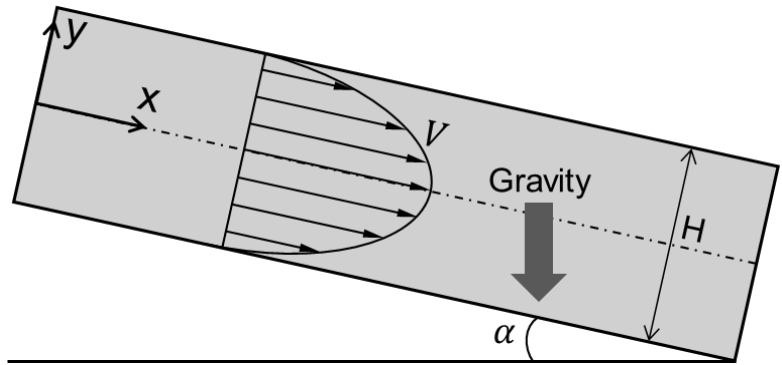

Figure 1 Schematic of the inclined plates.

The vectorial (expanded) form of the governing equation for the conservation of linear momentum is:

$$\phi\rho\left[\frac{\partial \boldsymbol{V}}{\partial \tau} + (grad\boldsymbol{V})\boldsymbol{V}\right]$$
$$= (grad\phi)B_p + \phi grad B_p + (grad\phi^2)h(\phi)B_r + \phi^2 grad(h(\phi)B_r)$$
$$+ [B_{31}(\phi + \phi^2) + B_{32}(\phi + \phi^2)\Pi]div\boldsymbol{A}_1 + B_{31}\boldsymbol{A}_1 grad(\phi + \phi^2) \quad (19)$$
$$+ B_{32}\boldsymbol{A}_1[\Pi grad(\phi + \phi^2) + (\phi + \phi^2)grad\Pi] + \frac{\rho}{Fr}\phi\boldsymbol{b}$$
$$+ M_1 div\big((\phi + \phi^2)\boldsymbol{A}_2\big) + M_2 div\big((\phi + \phi^2)\boldsymbol{A}_1^2\big)$$

As a results of the non-dimensionalization, we obtain the following non-dimensional parameters:

$$Y = \frac{y}{H};\ X = \frac{x}{H};\ \boldsymbol{V} = \frac{\boldsymbol{v}}{u_0};\ \tau = \frac{tu_0}{H};\ \boldsymbol{b}^* = \frac{\boldsymbol{b}}{g};\ \rho^* = \frac{\rho}{\rho_0}$$

$$\text{div}^*(\cdot) = H\text{div}(\cdot);\ \text{grad}^*(\cdot) = H\text{grad}(\cdot);\ \boldsymbol{L}^* = \text{grad}^*\boldsymbol{V}$$

$$\boldsymbol{A}_1^* = \boldsymbol{L}^* + \boldsymbol{L}^{*T};\ \boldsymbol{A}_2^* = \frac{d\boldsymbol{A}_1^*}{d\tau} + \boldsymbol{A}_1^*\boldsymbol{L}^* + \boldsymbol{L}^{*T}\boldsymbol{A}_1^*$$

$$Fr = \frac{u_0^2}{Hg};\ B_p = \frac{\beta_p}{\rho_0 u_0^2};\ B_r = \frac{C_r}{\rho_0 u_0^2};\ B_{31} = \frac{\mu_\infty}{\rho_0 u_0 H};\ B_{32} = \frac{\mu_0 - \mu_\infty}{\rho_0 u_0 H}; \quad (20)$$

$$M_1 = \frac{\alpha_{10}}{\rho_0 H^2};\ M_2 = \frac{\alpha_{20}}{\rho_0 H^2};\ \Pi = \frac{1 + \ln(1 + k\dot{\gamma})}{1 + k\dot{\gamma}}$$

$$G = \frac{\rho}{Fr};\ \Gamma = \frac{H\dot{\gamma}}{u_0};\ \bar{k} = \frac{ku_0}{H};$$

where $H$ is a reference length, for example, the distance between the two plates, and $u_0$ is a reference velocity. It should be noticed that in equation (19) the asterisks have been dropped for simplicity. Among the above dimensionless numbers, $B_p$, $M_1$ and $M_2$ are related to the normal stress coefficients, $B_r$ is related to the compactness effect. $B_p$ and $B_r$ are always less than zero, implying that compression should lead to densification of the granular materials. $Fr$ is the Froude number, $\bar{k}$ is a parameter related to the shear-thinning effects, $B_{31}$ and $B_{32}$ are related to the viscous effects (similar to the Reynolds number).

Furthermore, we assume that the flow is steady and fully developed,

$$\boldsymbol{v} = U(Y)\boldsymbol{e}_x;\ \phi = \phi(y) = \rho(y)/\rho_0 \quad (21)$$

where $\boldsymbol{e}_x$ is the unit vector along the *x*-direction. With Equation (21) we have,

$$\boldsymbol{D} = \frac{1}{2}(grad\boldsymbol{v} + (grad\boldsymbol{v})^T) = \frac{1}{2}\begin{bmatrix} 0 & U' & 0 \\ U' & 0 & 0 \\ 0 & 0 & 0 \end{bmatrix} \quad (22)$$

With above equations, the equations for the conservation of mass is automatically satisfied. In other words, the granular materials are incompressible in the sense that,

$$tr\mathbf{D} = div\mathbf{v} = 0 \tag{23}$$

Substituting (21) and (23) into (19), the governing equations are simplified and we obtain the two coupled ordinary equations.

The momentum equation in the x-direction is:

$$2(\phi + \phi^2)\left[B_{31} + B_{32}\Pi + \left(B_{32}\frac{1}{\Gamma}\frac{d\Pi}{d\Gamma}U'^2\right)\right]U'' + 2(1 + 2\phi)\phi'(B_{31} + B_{32}\Pi)U' \\ + \phi G sin(\alpha) = 0 \tag{24}$$

The momentum equation in the y-direction is:

$$(B_p)\phi' + 2B_r\phi\phi'h(\phi) + B_r\phi^2 h'(\phi) + (2M_1 + M_2)\phi'U'^2(1 + 2\phi) \\ + 2(2M_1 + M_2)(\phi + \phi^2)U'U'' - \phi G cos(\alpha) = 0 \tag{25}$$

where $U$ is the velocity and here we assume $b = g$, thus $b^* = 1$. It is worth pointing out that the normal stress difference for this problem are,

$$T_{xx} - T_{yy} = -2M_1(\phi + \phi^2)U'^2 \\ T_{yy} - T_{zz} = 2M_1(\phi + \phi^2)U'^2 + M_2(\phi + \phi^2)U'^2 \tag{26}$$

Ordinary differential equations (24) and (25) need to be solved numerically. From these equations, we can see that we need two boundary conditions for $U$ and one boundary condition for $\phi$. In this paper, we use the no-slip velocity condition at the both boundaries:

$$U(Y = 1) = U(Y = -1) = 0 \tag{27}$$

For volume fraction, $\phi$, the appropriate boundary condition may be given as an average volume fraction given in an integral form:

$$\int_{-1}^{1} \phi \, dY = N \tag{28}$$

Alternatively, the value of $\phi$ could be given at $Y = -1$:

$$\phi \to \Theta \text{ as } Y \to -1 \tag{29}$$

where $\Theta$ is the value of the volume fraction at the boundary. In the next section, we perform a parametric study for a selected range of the dimensionless numbers.

## 5. Results and Discussions

In this paper, the system of the non-linear ordinary differential equations (24) and (25) with the boundary conditions (27) and (28) are solved numerically using the MATLAB solver bvp4c, which is a collocation boundary value problem solver[46]. The step size is automatically adjusted by the solver. The default relative tolerance for the maximum residue is 0.001. The boundary conditions for the average/bulk volume fraction is numerically satisfied by using the shooting method.

*5.1. The effect of the particle compactness*

First, we consider the effect of the newly proposed isotropic (spherical) part of the stress tensor due to the particle contact.

Figure 2 shows the effect of $B_r$ on the velocity and volume fraction profiles. The granular materials with 'harder' solid particles have larger $B_r$. From the figures, we can see that overall, due to gravity, more particles accumulate near the bottom plate and the particles concentration decreases from the bottom plate to top plate; as a result of the particles distribution, the location of the maximum velocity is closer to the top plate. The shift in the location of the maximum velocity can be attributed to a dense particle region causing more resistance to the flow. From the volume fraction profiles, we can see that as $B_r$ increases, the volume fraction near the bottom plate decreases; when $B_r$ is large, for example $B_r = 10$, the maximum volume fraction near the bottom plate is 0.68 which is the value of $\phi_c$. We may infer that for granular materials with very rigid particles, the value of $B_r$ should be large, ensuring that the volume fraction is always bounded the value of the critical volume fraction $\phi_c$. As $B_r$ increases, the value of the maximum velocity increases. [Recall that in this paper, the flow is driven by the gravity. Thus, we consider any change in the velocity is related to the non-uniform distribution of the particles. For a pressure driven flow, the situation may change.]

Figure 3 shows the effect of another important parameter, namely $\phi_c$, which represents the critical volume fraction related to the maximum particle packing, close or above which the granular material shows the spherical isotropic stress, $\beta_r$, preventing further compacting. As $\phi_c$ changes, the variation in the flow field becomes significant, especially for the volume fraction profiles. When $\phi_c$ decreases, the volume fraction near the bottom plate changes dramatically. Furthermore, since the bulk volume fraction is a constant (assumed to be 0.4 in this paper), for larger values of $\phi_c$, the thickness of the sedimentation region near the bottom plate becomes smaller. As $\phi_c$ increases, the velocity decreases.

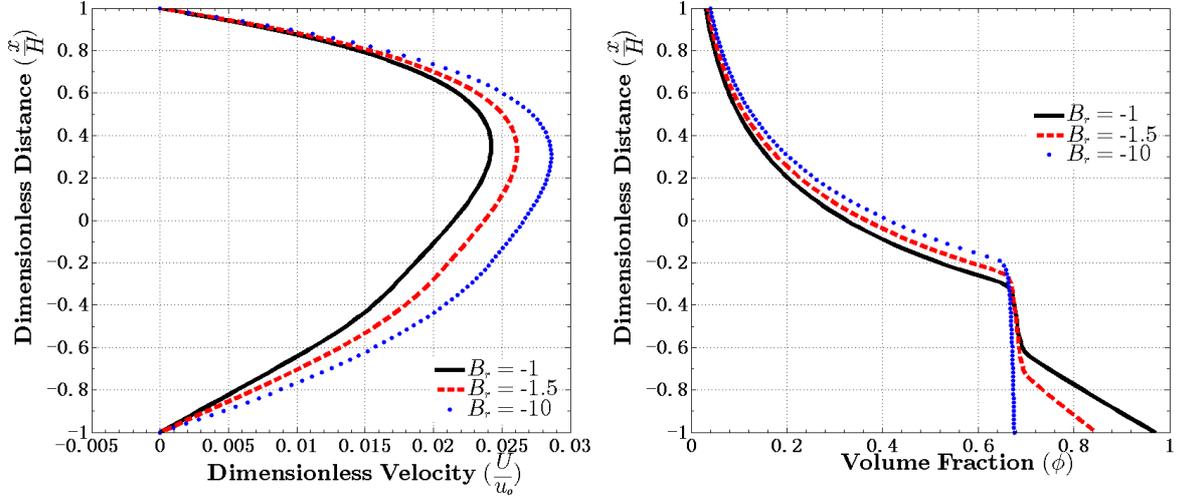

Figure 2 The effect of $B_r$ on the velocity profile (left) and the volume fraction profile (right). With $B_{31} = 1$, $B_{32} = 5$, $k = 10$, $(2M_1 + M_2) = 3$, $B_p = -1$, $\phi_c = 0.68$, $S = 100$, $\alpha = 20°$, $G = 2.5$, $N = 0.4$.

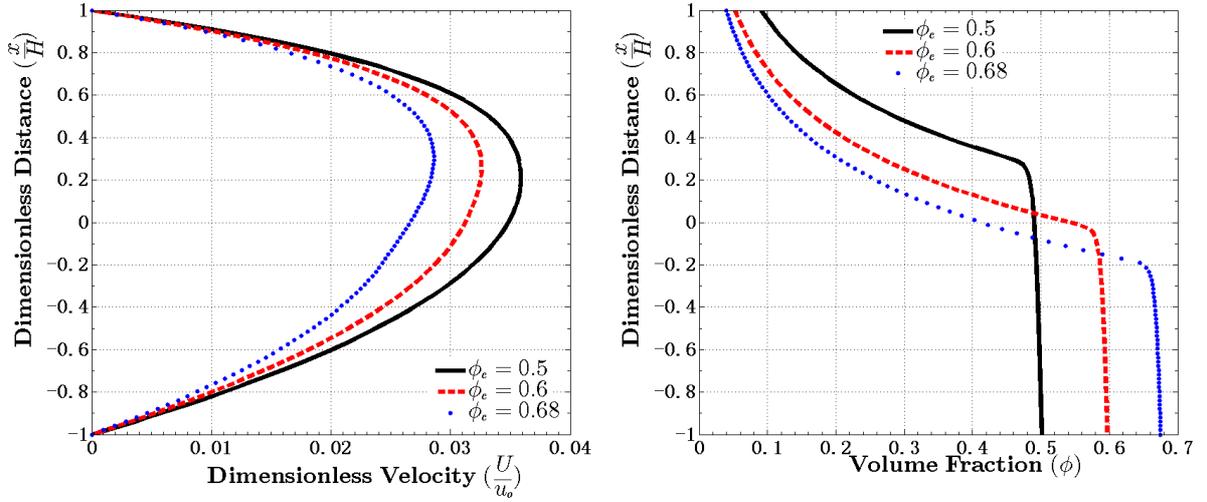

Figure 3 The effect of $\phi_c$ on the velocity profile (left) and the volume fraction profile (right). With $B_{31} = 1$, $B_{32} = 5$, $k = 10$, $(2M_1 + M_2) = 3$, $B_p = -1$, $B_r = -10$, $S = 100$, $\alpha = 20°$, $G = 2.5$, $N = 0.4$.

### 5.2. Normal stress differences and the shear-thinning effects of the viscosity

In our model, the normal stress differences are related to the parameters $M_1$ and $M_2$. As Figure 4 indicates, for the range of the parameters chosen, the effect of $(2M_1 + M_2)$ is not very significant. As $(2M_1 + M_2)$ increases, namely as the effect of the normal stresses become stronger, the volume fraction near the bottom plate does not change much. From equation (25), we can see that the volume fraction distribution (along the y-direction) is determined by the competition among the normal stresses $(2M_1 + M_2)$ and $B_p$, the gravity term $(G)$ and $B_r$. The small variation of the volume fraction profile near the bottom plate may imply that gravity dominates for the range of the dimensionless parameters considered in this case. We also see that near the top plate, when $(2M_1 + M_2)$ increases, more particles seem to move towards the top plate and therefore the volume fraction increases. Furthermore, as $(2M_1 + M_2)$ increases the velocity seems to decrease. Figure 5 shows the effect of the normal stress difference, when $\alpha = 80°$ (In Figure 4, $\alpha = 20°$.). As Figure 5 shows, the velocity profiles appear to be more symmetric near the centerline where the effect of $(2M_1 + M_2)$ is more obvious on both the velocity and the volume fraction profiles. As $(2M_1 + M_2)$ increases, more particles move towards the plates, and especially when $(2M_1 + M_2) = 100$, the volume fraction near the plate is approaching $\phi_c$ Increasing $(2M_1 + M_2)$ causes a decrease in the velocity, possibly due to the accumulation of particles near the plates.

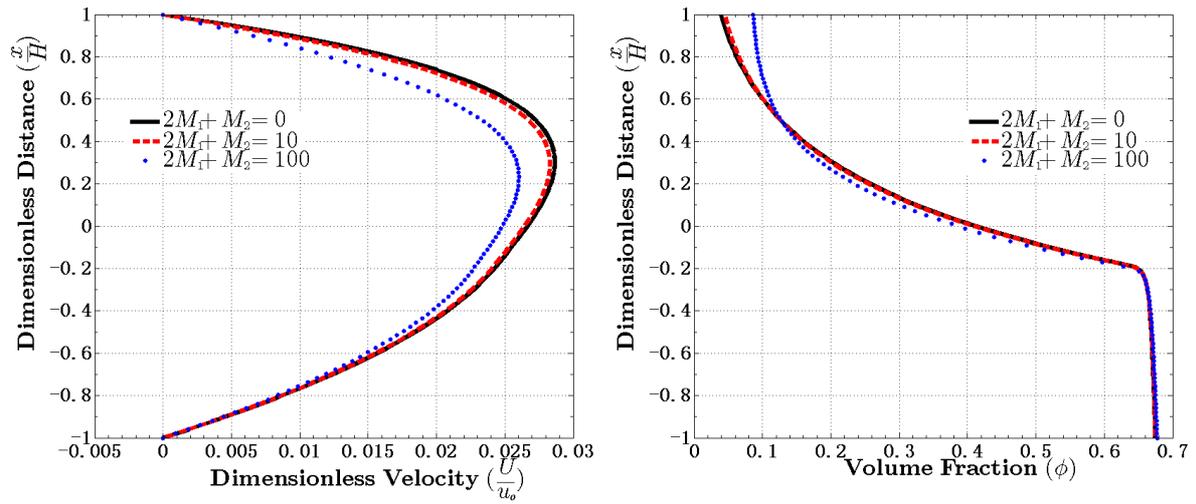

Figure 4 The effect of $(2M_1 + M_2)$ on the velocity profile (left) and the volume fraction profile (right). With $B_{31} = 1$, $B_{32} = 5$, $k = 10$, $B_p = -1$, $B_r = -10$, $\phi_c = 0.68$, $S = 100$, $\alpha = 20°$, $G = 2.5$, $N = 0.4$.

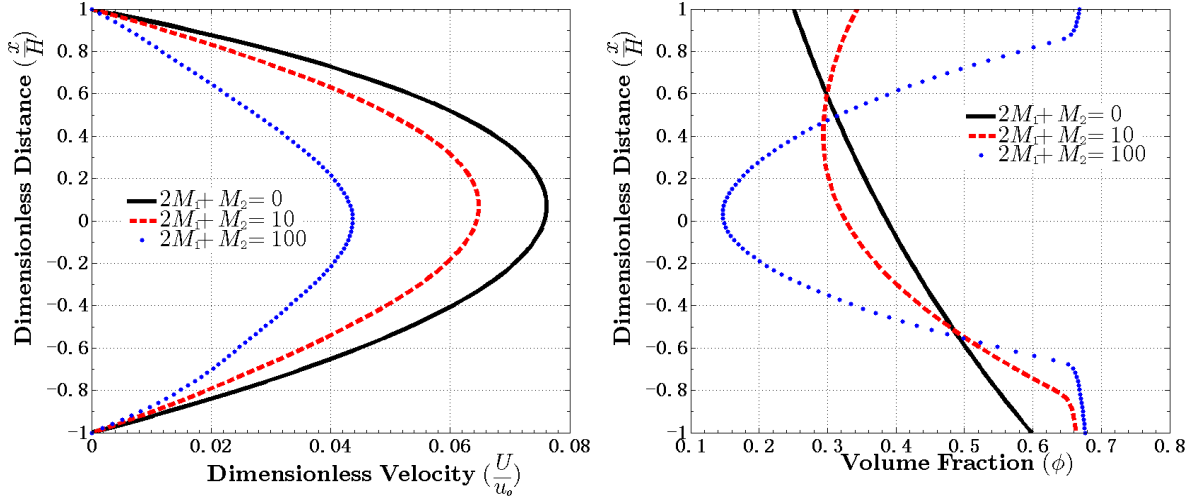

Figure 5 The effect of $(2M_1 + M_2)$ on the velocity profile (left) and the volume fraction profile (right). With $B_{31} = 1$, $B_{32} = 5$, $k = 10$, $B_p = -1$, $B_r = -10$, $\phi_c = 0.68$, $S = 100$, $\alpha = 80°$, $G = 2.5$, $N = 0.4$.

Figure 6(left) indicates that when $B_{31}$ and $B_{32}$ are small, namely when the viscosity is small, the values of the velocity are larger than the case with higher viscosities (see the cases when $B_{31} = 1$, $B_{32} = 1$, and $B_{31} = 1.5$, $B_{32} = 20$). As Figure 6(right) shows, when $B_{31}$ and $B_{32}$ increase, the volume fraction near the bottom plate decreases. If we look at the volume fraction curve when $B_{31} = 1.5$ and $B_{32} = 20$, we can see that the maximum volume fraction is smaller than $\phi_c$, which implies fewer particles have settled down. Near the top plate, when $B_{31}$ and $B_{32}$ are small, the particle concentration is higher the closer we get to the top plate.

Figure 7 shows the effect of $k$, the parameter related to the shear-thinning property. When $k = 0$, the viscosity reduces to $\mu_0$; when $k \to \infty$, the viscosity reduces to $\mu_\infty$. As Figure 7 shows, the velocity decreases as $k$ decreases, since a decrease in $k$ implies an increase in the viscosity. Accordingly, we can see that as $k$ increases, the volume fraction profiles become more non-uniform.

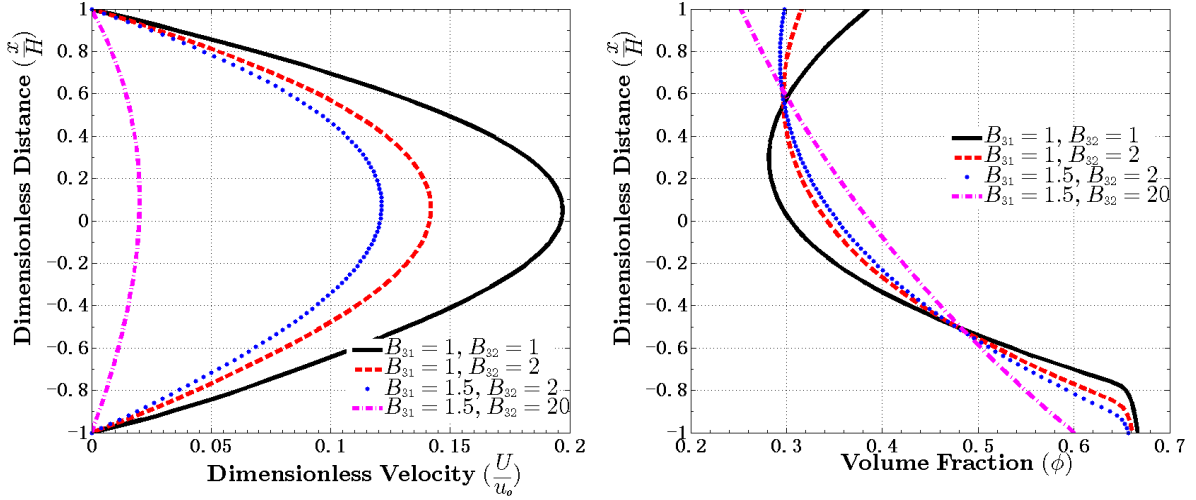

Figure 6 The effect of $B_{31}$ and $B_{32}$ on the velocity profile (left) and the volume fraction profile (right). With $k = 10, Re = 1, (2M_1 + M_2) = 3, B_p = -1, B_r = -10, \phi_c = 0.68, S = 100, \alpha = 80°, G = 2.5, N = 0.4$.

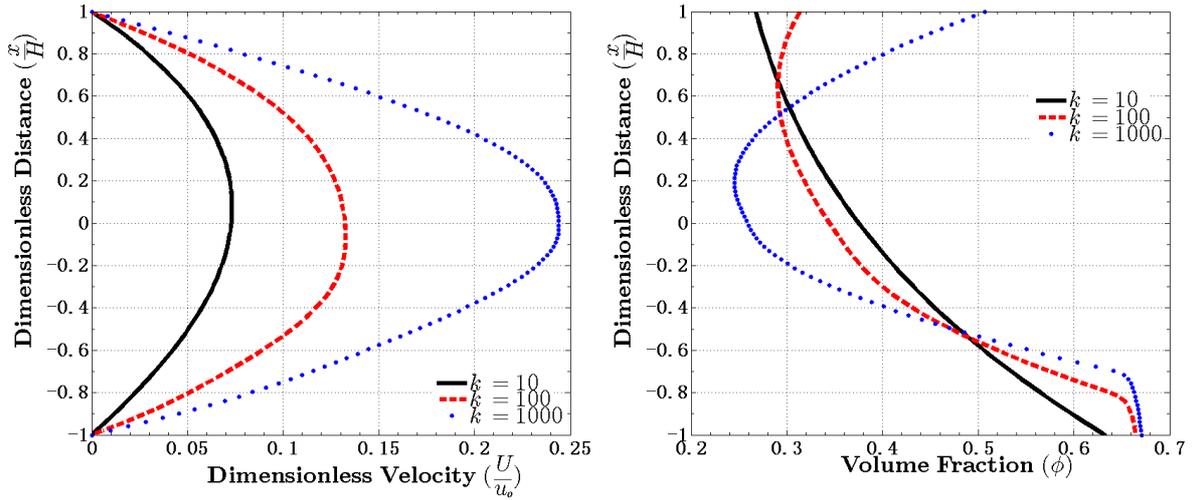

Figure 7 The effect of k on the velocity profile (left) and the volume fraction profile (right). With $B_{31} = 1, B_{32} = 5, (2M_1 + M_2) = 3, B_p = -1, B_r = -10, \phi_c = 0.68, S = 100, \alpha = 80°, G = 2.5, N = 0.4$.

### 5.3. Gravity and bulk volume fraction

Next we consider the effect of the parameters related to the inclination angle ($\alpha$), gravity ($G$), and the bulk volume fraction ($N$). Figure 8(left) shows that as $\alpha$ increases, the velocity increases and the maximum velocity moves towards the lower plate. From Figure 8(right), we can see that as $\alpha$ decreases, more and more particles accumulate near the lower plate, and a sedimentation region with a high concentration of particles is formed; here the volume fraction is close to $\phi_c$. Figure 9(left) shows that when $G$ increases (i.e. particles with larger density, $\rho$), the velocity increases and the position of the maximum velocity shifts towards the upper plate. From Figure 9(right), we can see that the thickness of the sedimentation region becomes larger as $G$ increases. Figure 10 shows the effect of the bulk volume fraction ($N$) on the velocity and volume fraction. As $N$ increases, more particles accumulate near the bottom plate, and the velocity becomes larger.

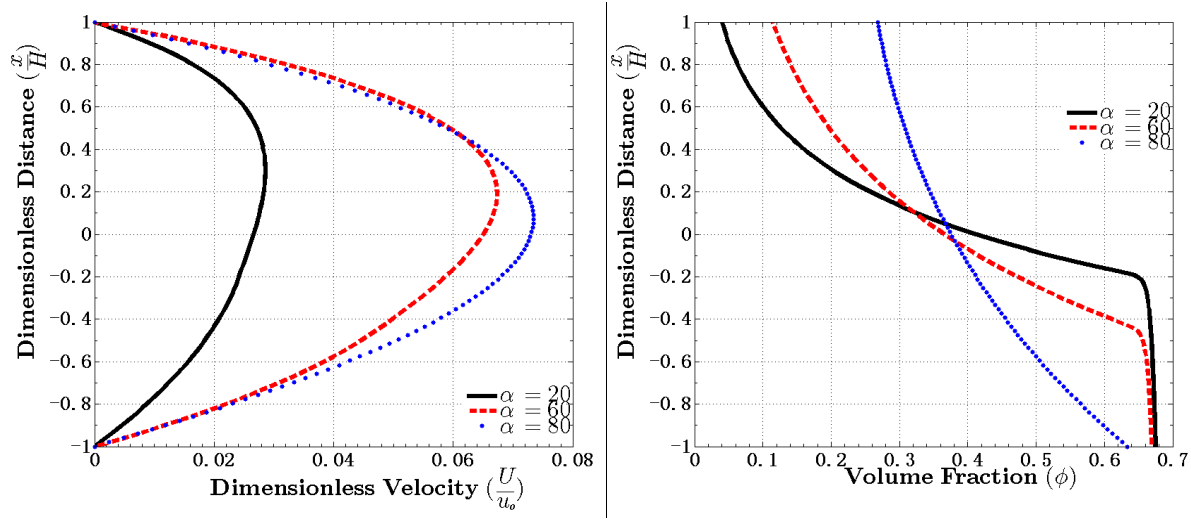

Figure 8 The effect of $\alpha$ on the velocity profile (left) and the volume fraction profile (right). With $B_{31} = 1, B_{32} = 5, k = 10, (2M_1 + M_2) = 3, B_p = -1, B_r = -10, \phi_c = 0.68, S = 100, G = 2.5, N = 0.4.$

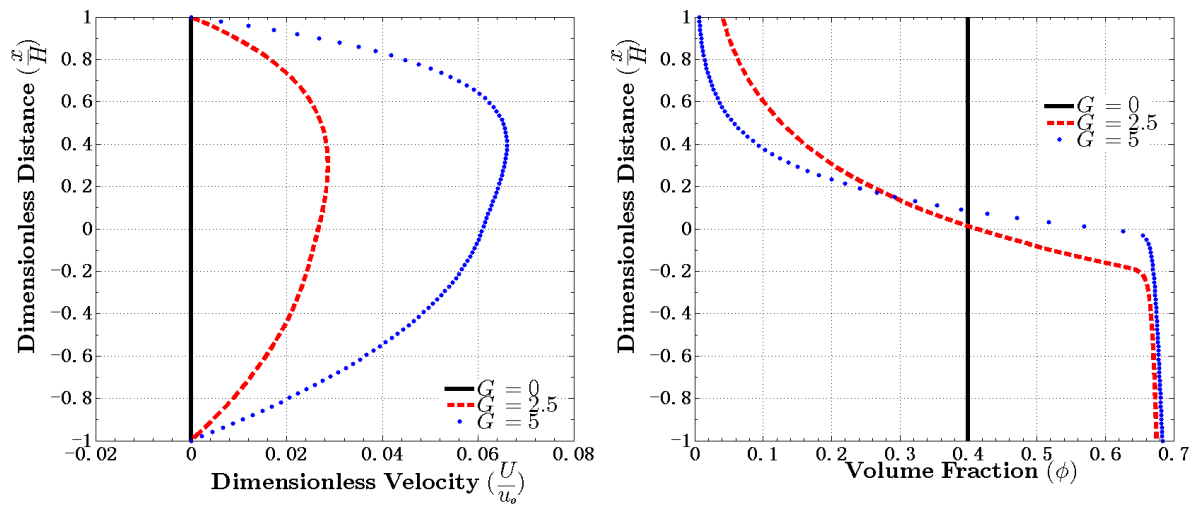

Figure 9 The effect of $G$ on the velocity profile (left) and the volume fraction profile (right). With $B_{31} = 1, B_{32} = 5, k = 10, (2M_1 + M_2) = 3, B_p = -1, B_r = -10, \phi_c = 0.68, S = 100, \alpha = 20°, N = 0.4.$

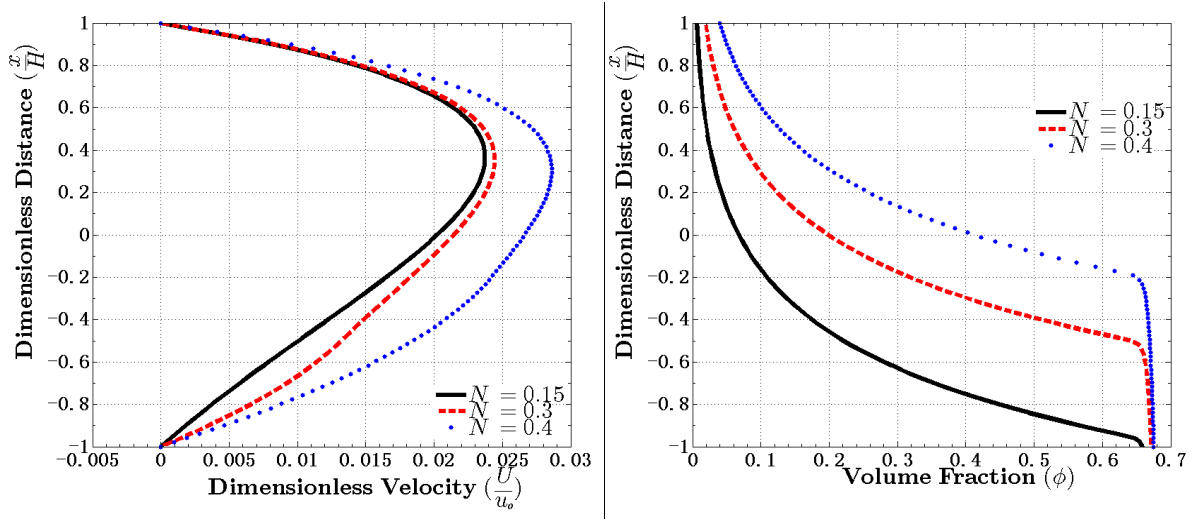

Figure 10 The effect of N on the velocity profile (left) and the volume fraction profile (right). With $B_{31} = 1$, $B_{32} = 5$, $k = 10$, $(2M_1 + M_2 = 3)$, $B_p = -1$, $B_r = -10$, $\phi_c = 0.68$, $S = 100$, $\alpha = 20°$, $G = 2.5$.

## 6. Conclusions

In this paper, we study the gravity driven fully developed flow of granular materials between two inclined plates. The granular materials are modeled as a non-linear fluid, which includes the effects of shear and volume fraction dependent viscosity and normal stress difference effects. We also propose a new term in the spherical part of the stress tensor, related to the compactness of the particles. This term is related to the critical volume fraction of the particles ($\phi_c$). This new term includes a smooth step function, and ensures that the new term only becomes effective when the local volume fraction is approaching $\phi_c$. A parametric study is performed for a selected range of dimensionless numbers.